\documentclass[conference]{IEEEtran}

\usepackage[english]{babel}
\usepackage[utf8]{inputenc}
\usepackage{amsmath}
\usepackage{graphicx}
\usepackage{url}
\usepackage[colorinlistoftodos]{todonotes}
\usepackage[normalem]{ulem}
\usepackage{float}
\usepackage{subfig}
\usepackage{supertabular}
\usepackage{multicol}
\usepackage{endnotes}

\title{The Karlskrona manifesto for sustainability design}

\begin{document}

\maketitle






\noindent \textbf{\large{Introduction}} \hspace{3.25cm} \emph{Version 1.0, May 2015}\\

\vspace{-0.2cm} \noindent As software practitioners and researchers, we are part of the group of people who design the software systems that run our world. Our work has made us increasingly aware of the impact of these systems and the responsibility that comes with our role, at a time when information and communication technologies are shaping the future. We struggle to reconcile our concern for planet Earth and its societies with the work that we do. Through this work we have come to understand that we need to redefine the narrative on sustainability and the role it plays in our profession.\\

\vspace{-0.2cm} \noindent What is sustainability, really?  We often define it too narrowly. Sustainability is at its heart a systemic concept  and has to be understood on a set of dimensions, including social, individual, environmental, economic, and technical\endnote{\textbf{What are those dimensions, really?} \begin{itemize}                    
\item \textit{Individual sustainability} refers to maintaining individual human capital (e.g., health, education, skills, knowledge, leadership, and access to services).                
\item \textit{Social sustainability} aims at preserving the societal communities in their solidarity and services.
\item \textit{Economic sustainability} aims at maintaining capital and added value.
\item \textit{Environmental sustainability} refers to improving human welfare by protecting the natural resources: water, land, air, minerals and ecosystem services.
\item \textit{Technical sustainability} refers to longevity of information, systems, and infrastructure and their adequate evolution with changing surrounding conditions.    
\end{itemize}
}.\\

\vspace{-0.2cm} \noindent Sustainability is fundamental to our society. The current state of our world is unsustainable in more ways that we often recognize. Technology is part of the dilemma and part of possible responses. We often talk about the immediate impact of technology, but rarely acknowledge its indirect and systemic effects. These effects play out across all dimensions of sustainability over the short, medium and long term.\\

\vspace{-0.2cm} \noindent Software in particular plays a central role in sustainability. It can push us towards growing consumption of resources, growing inequality in society, and lack of individual self-worth. But it can also create communities and enable thriving of individual freedom, democratic processes, and resource conservation. As designers of software technology, we are responsible for the long-term consequences of our designs. Design is the process of understanding the world and articulating an alternative conception on how it should be shaped, according to the designer’s intentions. Through design, we cause change and shape our environment. If we don’t take sustainability into account when designing, no matter in which domain and for what purpose, we miss the opportunity to cause positive change. \\

\vspace{-0.2cm} \noindent \textbf{We recognize that} 
there is a rapidly increasing awareness of the fundamental need and desire for a more sustainable world, and
a lot of genuine desire and goodwill  - but this alone can be ineffective unless we come to understand that:\\

\begin{footnotesize}
\vspace{-0.05cm}
\noindent \textbf{There is} a narrow perception of sustainability that frames it as protecting the environment or being able to maintain a business activity. \\ \textbf{Whereas} as a systemic property, sustainability does not apply simply to the system we are designing, but most importantly to the environmental, economic, individual, technical and social contexts of that system, and the relationships between them. \\

\vspace{-0.05cm} 
\noindent \textbf{There is} a perception that sustainability is a distinct discipline of research and practice with a few defined connections to software. \\ \textbf{Whereas} sustainability is a pervasive concern that translates into discipline-specific questions in each area it applies. \\

 \vspace{-0.05cm} 
 \noindent \textbf{There is} a perception that sustainability is a problem that can be solved, and that our aim is to find the `one thing` that will save the world. \\ \textbf{Whereas} it is a `wicked problem` - a dilemma to respond to intelligently and learn in the process of doing so; a challenge to be addressed, not a problem to be solved.\\

\vspace{-0.05cm} 
\noindent \textbf{There is} a perception that there is a tradeoff to be made between present needs and future needs, reinforced by a common definition of sustainable development, and hence that sustainability requires sacrifices in the present for the sake of future generations. \\ \textbf{Whereas} it is possible to prosper on this planet while simultaneously improving the prospects for prosperity of future generations.\\

\vspace{-0.05cm} 
\noindent \textbf{There is} a tendency to focus on the immediate impacts of any new technology, in terms of its functionality and how it is used. \\ \textbf{Whereas} the following orders of effects have to be distinguished:
\textit{Direct, first order effects} are the immediate opportunities and effects created by the physical existence of software technology and the processes involved in its design and production.
\textit{Indirect, second order effects} are the opportunities and effects arising from the application and usage of software.
\textit{Systemic, third order effects}, finally, are the effects and opportunities that are caused by wide-scale use of software systems over time.\\

\vspace{-0.05cm} 
 \noindent \textbf{There is} a tendency to overly discount the future.
 The far future is discounted so much that it is considered for free (or worthless). Discount rates mean that long-term impacts matter far less than current costs and benefits. \\ \textbf{Whereas} the consequences of our actions play out over multiple timescales, and the cumulative impacts may be irreversible. \\

\vspace{-0.05cm} 
 \noindent \textbf{There is} a  tendency to think that taking small steps towards sustainability is sufficient, appropriate, and acceptable. \\ \textbf{Whereas} incremental approaches can end up reinforcing existing behaviours and lure us into a false sense of security. However, current society is 
so far from sustainability that deeper transformative changes are needed. \\

\vspace{-0.05cm} 
 \noindent \textbf{There is} a tendency to treat sustainability as a desirable quality of the system that should be considered once other priorities have been established.\\ \textbf{Whereas} is not in competition with a specific set of quality attributes against which it has to be balanced - it is a fundamental precondition for the continued existence of the system and influences many of the goals to be considered in systems design. \\
 
 \vspace{-0.05cm} 
 \noindent \textbf{There is} a desire to identify a distinct completion point to a given project, so
success can be measured at that point, with respect to
pre-ordained
criteria. \\ \textbf{Whereas} measuring success at one point in time fails to capture the effects that play out over multiple timescales, and so tells us nothing about long-term success. Criteria for success change over time as we experience those impacts.\\

\vspace{-0.05cm} 
 \noindent \textbf{There is} a narrow conception of the roles of system designers, developers, users, owners, and regulators and their responsibilities, and there is a lack of agency of these actors in how they can fulfill these responsibilities. \\ \textbf{Whereas} sustainability imposes a distinct responsibility on each one of us, and that responsibility comes with a right to know the system design and its status, so that each participant is able to influence the outcome of the technology application in both design and use. \\

\vspace{-0.05cm} 
 \noindent \textbf{There is} a tendency to interpret the codes of ethics for software professionals narrowly to refer to avoiding immediate harm to individuals and property. \\ \textbf{Whereas} it is our responsibility to address the potential harm from the 2nd and 3rd-order effects of the systems we design as part of our design process, even if these are not readily quantifiable.\newpage
 \end{footnotesize}

\noindent As a result, even though the importance of sustainability is increasingly recognized, many software systems are unsustainable, and the broader impacts of most software systems on sustainability are unknown.\\

\noindent \textbf{\large{Thus, we propose the following initial set of principles and commitments:}}\\

 \vspace{-0.05cm} \noindent \textbf{Sustainability is systemic.} Sustainability is never an isolated property.
Systems thinking has to be the starting point for the transdisciplinary common ground of sustainability.\\

 \vspace{-0.05cm} \noindent \textbf{Sustainability has multiple dimensions.} We have to include those dimensions into our analysis if we are to understand the nature of sustainability in any given situation. \\

 \vspace{-0.05cm} \noindent \textbf{Sustainability transcends multiple disciplines.}  Working in sustainability means working with people from across many disciplines, addressing the challenges from multiple perspectives.\\

 \vspace{-0.05cm} \noindent \textbf{Sustainability is a concern independent of the purpose of the system.} Sustainability has to be considered even if the primary focus of the system under design is not sustainability. \\

\vspace{-0.05cm} \noindent \textbf{Sustainability applies to both a system and its wider contexts}  There are at least two spheres to consider in system design: the sustainability of the system itself and how it affects sustainability of the wider system of which it will be part.\\
 
\vspace{-0.05cm} \noindent \textbf{Sustainability requires action on multiple levels.} Some interventions have more leverage on a system than others. Whenever we take action towards sustainability, we should consider opportunity costs: action at other levels may offer more effective forms of intervention.\\

\vspace{-0.05cm} \noindent \textbf{System visibility is a necessary precondition and enabler for sustainability design.} The status of the system and its context should be visible at different levels of abstraction and perspectives to enable participation and informed responsible choice.\\

\vspace{-0.05cm} \noindent \textbf{Sustainability requires long-term thinking.} We should assess benefits and impacts on multiple timescales, and include longer-term indicators in assessment and decisions.\\
 
 \vspace{-0.05cm} \noindent \textbf{It is possible to meet the needs of future generations without sacrificing the prosperity of the current generation.} Innovation in sustainability can play out as decoupling present and future needs. By moving away from the language of conflict and the trade-off mindset, we can identify and enact choices that benefit both present and future.\\

\vspace{-0.05cm} \noindent Sustainability design in the context of software systems is the process of designing systems with sustainability as a primary concern, based on a commitment to these principles.

\noindent \textbf{\large{So what now? How do we start?}}
\vspace{0.2cm}
\\Each of the following stakeholders can do something right now to get started.\\

\noindent \textbf{Software practitioners}:  Try to identify effects of your project on technical, economic, environmental sustainability. Start asking questions about how to incorporate the principles into daily practice. Think about the social and individual dimensions. Talk about it with your colleagues. \\

\vspace{-0.2cm} \noindent \textbf{Researchers}: Identify one research question in your field that can help us to better understand sustainability design. Discuss it with your peers and think about how sustainability impacts your research area. \\

\vspace{-0.2cm} \noindent \textbf{Professional associations}: Revise code of ethics and practice to incorporate principles and explicitly acknowledge the need to consider sustainability as part of professional practice.\\

\vspace{-0.2cm} \noindent \textbf{Educators}: Integrate sustainability design in curricula for software engineering and other disciplines and articulate competencies required for successful sustainability design.\\

\vspace{-0.2cm} \noindent \textbf{Customers}: Put the concern on the table. Demand it in the next project.\\

\vspace{-0.2cm} \noindent \textbf{Users}: Demand that the products you use demonstrate that their designers have considered all dimensions of sustainability.

\vspace{0.2cm}

\noindent \textbf{Signed,}

\noindent\textit{Christoph Becker}, University of Toronto \& Vienna University of Technology

\noindent \textit{Ruzanna Chitchyan}, University of Leicester

\noindent \textit{Leticia Duboc}, State University of Rio de Janeiro

\noindent \textit{Steve Easterbrook}, University of Toronto

\noindent \textit{Martin Mahaux}, University of Namur

\noindent \textit{Birgit Penzenstadler}, California State University Long Beach

\noindent \textit{Guillermo Rodriguez-Navas}, Malardalen University

\noindent \textit{Camille Salinesi}, Universite Paris 1

\noindent \textit{Norbert Seyff}, University of  Zurich

\noindent \textit{Colin C. Venters}, University of Huddersfield

\noindent \textit{Coral Calero}, University of Castilla-La Mancha

\noindent \textit{Sedef Akinli Kocak}, Ryerson University

\noindent \textit{Stefanie Betz}, Karlsruhe Institute of Technology
\\

\theendnotes

\end{document}